\documentclass[3p,times]{elsarticle}
\usepackage{ecrc}
\usepackage{graphicx}
\usepackage{subfigure}
\usepackage{gensymb}
\usepackage[figuresright]{rotating}
\usepackage{color}
\usepackage{amsmath}
\usepackage{longtable}

\volume{00}

\firstpage{1}

\journalname{Science Bulletin}

\runauth{Hu et al.}

\begin{document}

\begin{frontmatter}



\title{Optical Observations of LIGO Source GW 170817 by the Antarctic Survey Telescopes at Dome A, Antarctica}


\author{
Lei Hu$^{1,2,3}$,
Xuefeng Wu$^{1,2,3}$,
I. Andreoni$^{4,5,6}$,
Michael C. B. Ashley$^{7}$,
J. Cooke$^{4,5,8}$,
Xiangqun Cui$^{9,2}$,
Fujia Du$^{9}$,
Zigao Dai$^{10}$,
Bozhong Gu$^{9}$,
Yi Hu$^{11,2}$,
Haiping Lu$^{9}$,
Xiaoyan Li$^{9}$,
Zhengyang Li$^{9}$,
Ensi Liang$^{10}$,
Liangduan Liu$^{10}$,
Bin Ma$^{11,2}$,
Zhaohui Shang$^{12,11,2}$,
Tianrui Sun$^{1,2,13}$,
N. B. Suntzeff$^{14}$,
Charling Tao$^{15,16}$,
Syed A. Uddin$^{1,2,8}$,
Lifan Wang$^{14,1,2*}$,
Xiaofeng Wang$^{16}$,
Haikun Wen$^{9}$,
Di Xiao$^{10}$,
Jin Xu$^{9}$,
Ji Yang$^{1}$,
Shihai Yang$^{9}$,
Xiangyan Yuan$^{9,2}$,
Hongyan Zhou$^{17}$,
Hui Zhang$^{10}$,
Jilin Zhou$^{10}$,
Zonghong Zhu$^{18,19}$
}

\address{
$^{1}$Purple Mountain Observatory, Chinese Academy of Sciences, Nanjing 210008, China\\
$^{2}$Chinese Center for Antarctic Astronomy, Nanjing 210008, China\\
$^{3}$School of Astronomy and Space Sciences, University of Science and Technology of China, Hefei, 230029, China\\
$^{4}$Centre for Astrophysics and Supercomputing, Swinburne University of Technology, PO Box 218, H29, Hawthorn, VIC 3122, Australia\\
$^{5}$The Australian Research Council Centre of Excellence for Gravitational Wave Discovery (OzGrav)\\
$^{6}$Australian Astronomical Observatory, 105 Delhi Rd, North Ryde, NSW 2113, Australia\\
$^{7}$School of Physics, University of New South Wales, NSW 2052, Australia\\
$^{8}$The Australian Research Council Centre of Excellence for All-Sky Astrophysics (CAASTRO)\\
$^{9}$Nanjing Institute of Astronomical Optics and Technology, Nanjing 210042, China\\
$^{10}$School of Astronomy and Space Science and Key Laboratory of Modern Astronomy and Astrophysics in Ministry of Education, Nanjing University, Nanjing 210093, China\\
$^{11}$National Astronomical Observatories, Chinese Academy of Sciences, Beijing 100012, China.\\
$^{12}$Tianjin Normal University, Tianjin 300074, China\\
$^{13}$Shanghai Keylab for Astrophysics, Shanghai Normal University, Shanghai 200234, China\\
$^{14}$George P. and Cynthia Woods Mitchell Institute for Fundamental Physics \& Astronomy, Texas A. \& M. University, Department of Physics and Astronomy, 4242 TAMU, College Station, TX 77843, USA\\
$^{15}$Aix Marseille Univ, CNRS/IN2P3, CPPM, Marseille, France\\
$^{16}$Physics Department and Tsinghua Center for Astrophysics (THCA), Tsinghua University, Beijing, 100084, China\\
$^{17}$Polar Research Institute of China, 451 Jinqiao Rd, Shanghai 200136, China\\
$^{18}$Department of Astronomy, Beijing Normal University, Beijing 100875, China\\
$^{19}$School of Physics and Technology, Wuhan University, Wuhan 430072, China\\
$^*$lifanwang@gmail.com\\
}

\begin{abstract}
The LIGO detection of gravitational waves (GW) from merging black holes in 2015 marked the beginning of a new era in observational astronomy. The detection of an electromagnetic signal from a GW source is the critical next step to explore in detail the physics involved. The Antarctic Survey Telescopes (AST3), located at Dome A, Antarctica, is uniquely situated for rapid response time-domain astronomy with its continuous night-time coverage during the austral winter. We report optical observations of the GW source (GW~170817) in the nearby galaxy NGC 4993 using AST3. The data show a rapidly fading transient at around 1 day after the GW trigger, with the $i$-band magnitude declining from $17.23\pm0.13$ magnitude to $17.72\pm0.09$ magnitude in $\sim 1.8$ hour. The brightness and time evolution of the optical transient associated with GW~170817 are broadly consistent with the predictions of models involving merging binary neutron stars. We infer from our data that the merging process ejected about $\sim 10^{-2}$ solar mass of radioactive material at a speed of up to $30\%$ the speed of light.
\end{abstract}


\end{frontmatter}

\section{Introduction}
On August 17, 2017, the LIGO-Virgo gravitational wave (GW) detector network observed a GW signal from a binary neutron star (BNS) merger, referred to as GW 170817 \citep{lvc17a,lvc17b,lvc17c,lvc17d,lvc17e,lvc17f}. A short gamma ray burst (SGRB), GRB 170817A, was detected by the GBM instrument on board NASA's \emph{Fermi} satellite less than 2 seconds after the GW detection \citep{connaughton17}. The object was also observed by ESA's \emph{INTEGRAL} satellite \citep{savchenko17}. The association of a $\gamma$-ray burst with GW 170817A triggered a massive number of follow-up observations by ground-based telescopes which very quickly led to the identification of the optical counterpart of GW 170817 in a nearby galaxy NGC 4993. This intensive campaign provides strong evidence for the BNS merger scenario of GW 170817 and marks the beginning of a new era of multi-messenger astronomy.

Following the announcement of GW 170817 and GRB 170817A, the optical counterpart known as AT2017gfo was discovered by the One-Meter, Two Hemisphere (1M2H) team using the 1-m Swope telescope. The object was found to be located in the galaxy NGC 4993, and the first detection was carried out less than 11 hours after the detection of GW 170817 \citep{coulter17}. Many other teams also made independent discoveries and observations of this target \citep{lvc17g}.

BNS mergers produce rapidly evolving optical and infrared transients accompanied by the radiation of gravitational waves \citep{barn13,tana14,kasen15,metz15,barn16}. Radioactive materials synthesized in the rapid neutron capture process (r-process) are ejected during the merger. The decay of these radioactive materials results in optical and infrared emission with a typical duration of a day to a week \citep{li98,kul05,met10,kas13,bar13,tan13,bar16,met17}, and a peak luminosity that is about a few thousand times that of a typical nova. Such transient objects in the optical are given the name kilonovae (or macronovae). In contrast to binary black hole mergers, where there is no consensus on the detectability of electromagnetic radiation after the merger, kilonovae are expected to be detectable in both GW and optical/IR bands at a distance up to a few 10 Mpc. Before GW 170817, several kilonovae candidates were identified during follow-up observations of SGRBs, including GRB 130603B \citep{tani13,ber13}, GRB 050709 \citep{jin16}, GRB 060614 \citep{yan15}, and GRB 080503 \citep{perley09,gao15}. Simultaneous detection of a kilonova and a SGRB associated with a GW event would provide key clues to the properties of the ejecta of BNS mergers.

In this paper, we present follow up observations of GW 170817 and its electromagnetic counterpart AT2017gfo by AST3-2, the second telescope of the Antarctic Survey Telescopes at Dome A, Antarctica. In Section 2, we present the instrument, observations, and data reduction in detail. In Section 3, we apply an analytic kilonova model to the temporal evolution of the optical transient associated with GW 170817 acquired by AST3-2 to constrain the most fundamental physical properties related to GW 170817. Our data establishes that optical signals from GW 170817 are consistent with kilonova models. We summarize our results in Section 4.

\section{Instrument, Observations and Data Reduction}
The AST3-2 is a catadioptric optical telescope with an entrance pupil diameter of 500 mm and an f-ratio of 3.73. The telescope is located at Dome A, Antarctica \citep{burt10,saun09}. Its unique location allows for continuous observations lasting longer than 24 hours during the austral winter. The continuous time coverage is important for many applications of time-domain astronomy (e.g., \cite{wang13}, \cite{wang17} and \cite{yang17}).
It is equipped with a 10K$\times$10K CCD camera in frame transfer mode that provides a field of view of $4.14$ deg$^{2}$ \citep{yuan12}. 
The AST3 camera has 16 amplifiers with half of the area masked for frame transfer charge storage. The camera does not need a mechanical shutter to operate. The entire observatory is controlled remotely using the Iridium satellite system to transmit operational commands. Dome A is a site with the lowest temperature and the lowest absolute humidity on earth\citep{yang17,hu14,zhou13,wange17}. The weather conditions are suitable for taking data for over 95\% of the time during the winter. We achieved a full year of remotely controlled operation for the first time during the austral winter in 2017.


The optical counterpart AT2017gfo is located at the coordinates RA\,$=13$:09:48.089 and DEC\,$=-23$:22:53.35, about 2.2\,kpc from the center of the host galaxy NGC 4993\citep{coulter17}. AST3-2 observed this target from August 18 through August 28 in $i$-band\citep{hu17}. The coordinates of the target are not ideal for AST3-2, which is located at a latitude of $-80^\circ\,22^\prime$. At the time of observations, the maximum elevation of the Sun was already too high to allow for multi-day continuous observations. Nevertheless, a total of 262 exposures were obtained, with each having an exposure time of 300\,s except for the first 5 images with exposure times of 60\,s. The gap between the exposures is typically 54\,s. Due to the low altitude of the target and the rising Sun, the target was only observable for about 2 hours each day. Data reported in this paper consist of a total of 91 images taken during the period from Aug 18, 2017 to Aug 23, 2017. Data taken after Aug. 24, 2017 show no detectable transients at the position of GW 170817 and will not be discussed in this paper. Data adjacent in time are stacked to produce the final light curves, the dates of observations are shown in Table~1.

\begin{figure*}
\centering
\includegraphics[scale=0.5]{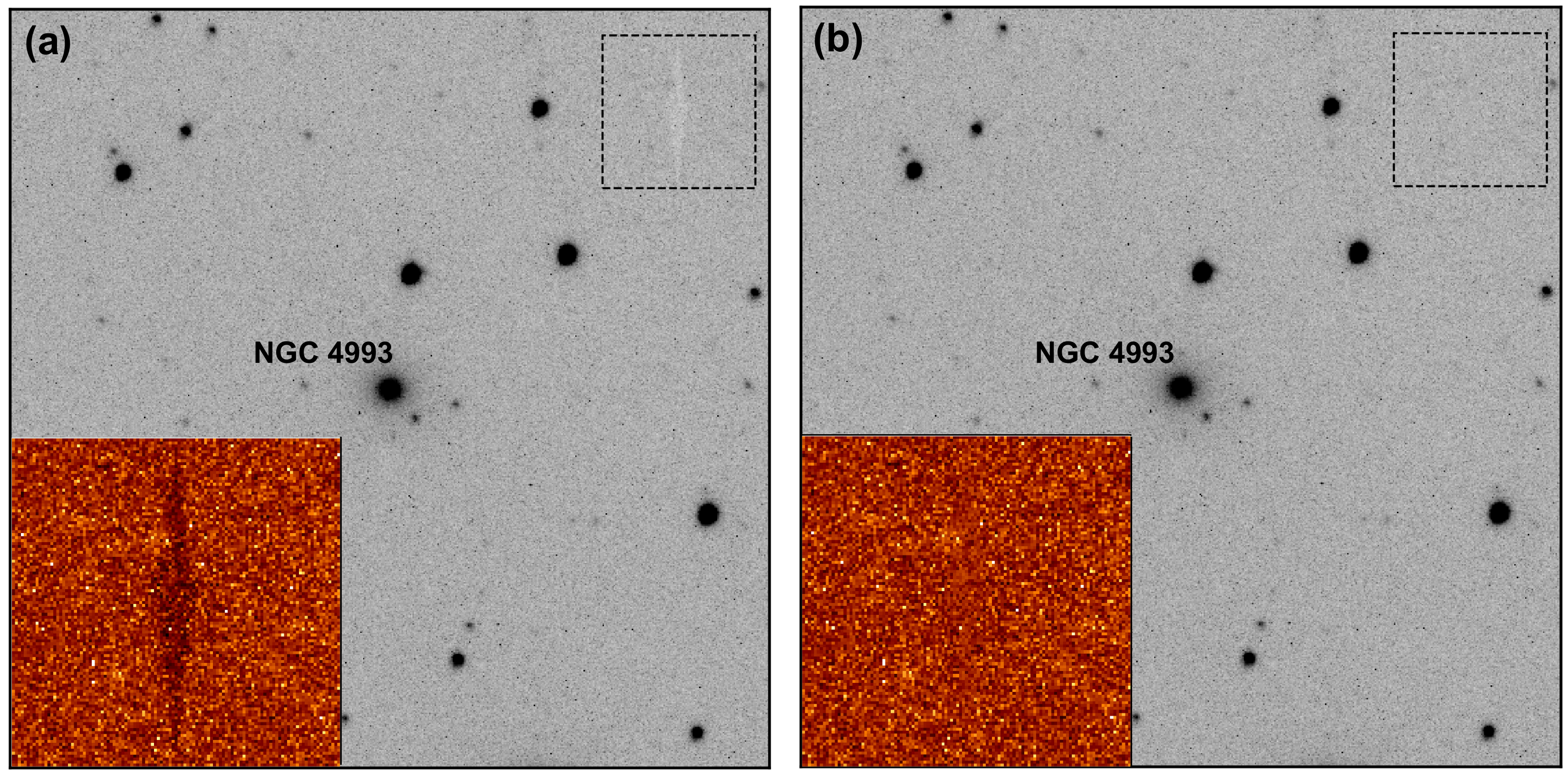}
\caption{(a) Raw image with cross-talk contamination uncorrected. A dashed square in the upper right corner marks the position of a typical cross-talk affected region. The details of the region is shown in the insert at the lower left corner. (b) Same as (a), but with the cross-talk affected region corrected.}
\label{fig1}
\end{figure*}

Data reduction follows the standard procedure of bias and overscan subtraction, and flat-field correction. Cosmic rays were removed with a Laplacian Edge Detection algorithm, which is a reliable method for identifying and removing cosmic rays\citep{van01}. Special treatment was applied to the removal of the structures produced by saturated stars that spread to all the 16 readout channels through cross-talks among the channels. 
It is possible to construct a very accurate correction of the effect of cross-talk. The pixels affected by cross-talks follow a well defined pattern. The strength of the cross-talk is a function of the pixel values of the (near) saturated pixels \citep{Frey01} of the stars causing the cross-talk. 
Figure \ref{fig1} shows an image affected by amplifier cross-talk and the same image after correction. 


The relatively bright host galaxy makes direct aperture photometry difficult.  Difference images using late time observations as a template were needed to be constructed for reliable photometry. This required precise astrometric registration of the images. 
After the preliminary overscan and flat-field corrections, all sky background subtracted images were re-aligned by matching the coordinates of field stars in all images to their corresponding coordinates in the reference frame. The reference frame was constructed using an image taken on Aug 23, 2017 in the best seeing conditions. During the matching process, we employed a Lanczos-3 interpolation kernel proposed by \citet{bert02}. The weight of each registered image during the co-adding process was derived from the equation recommended by \citep{ann14}, which was successfully applied to the co-adding of SDSS Stripe 82 images,
\begin{equation}
w = \frac{F}{\hbox{\rm FWHM}^2\,V}
\end{equation}
where $F$ is characterized by the flux-based photometric zero point, and $V$ is the variance of background noise. Such a weighting scheme gives larger weights to images with good seeing conditions and low background levels.



\begin{figure*}
\centering
\subfigure{
\centering
\includegraphics[scale=0.5]{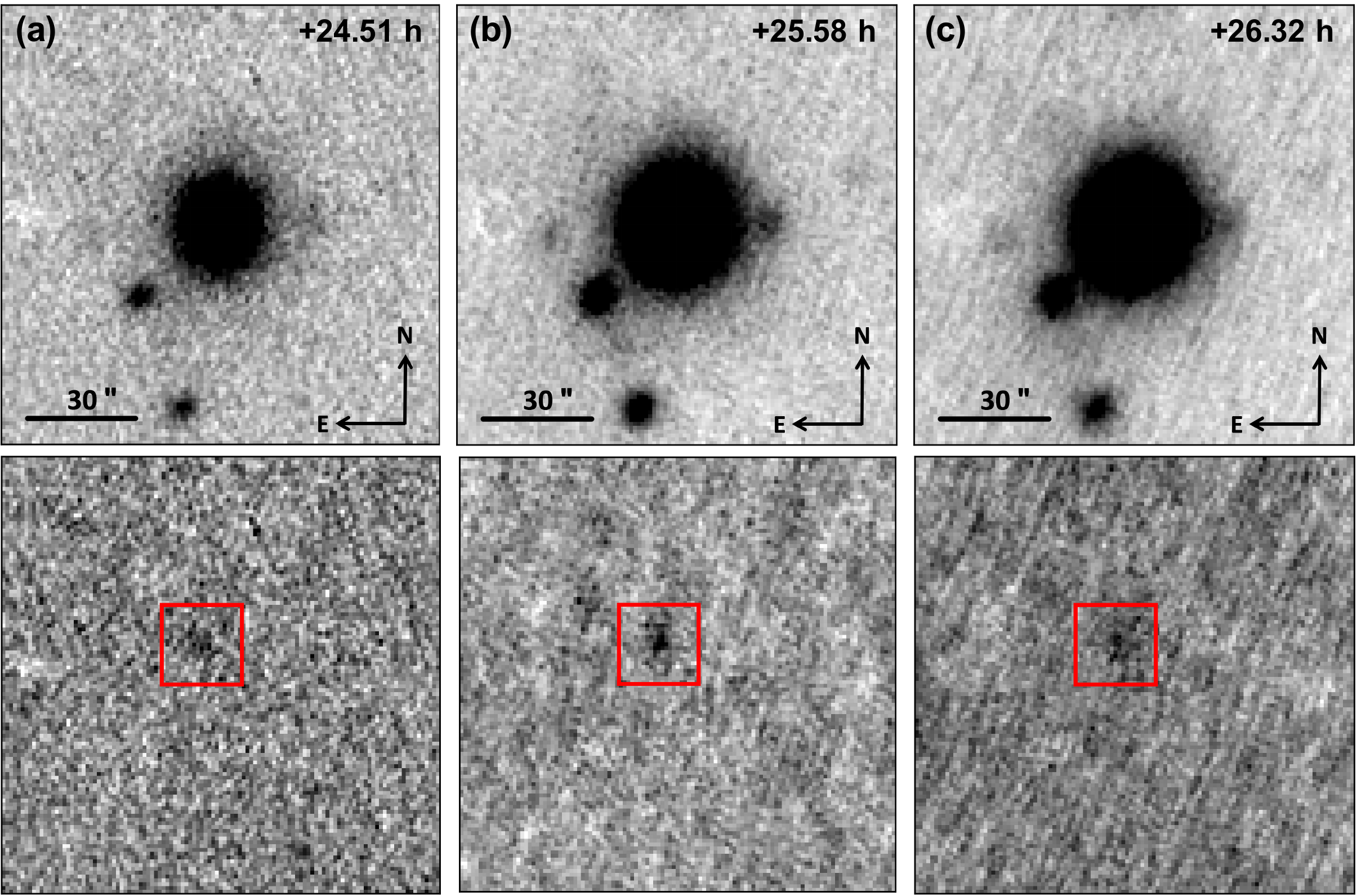}
}

\subfigure{
\centering
\includegraphics[scale=0.5]{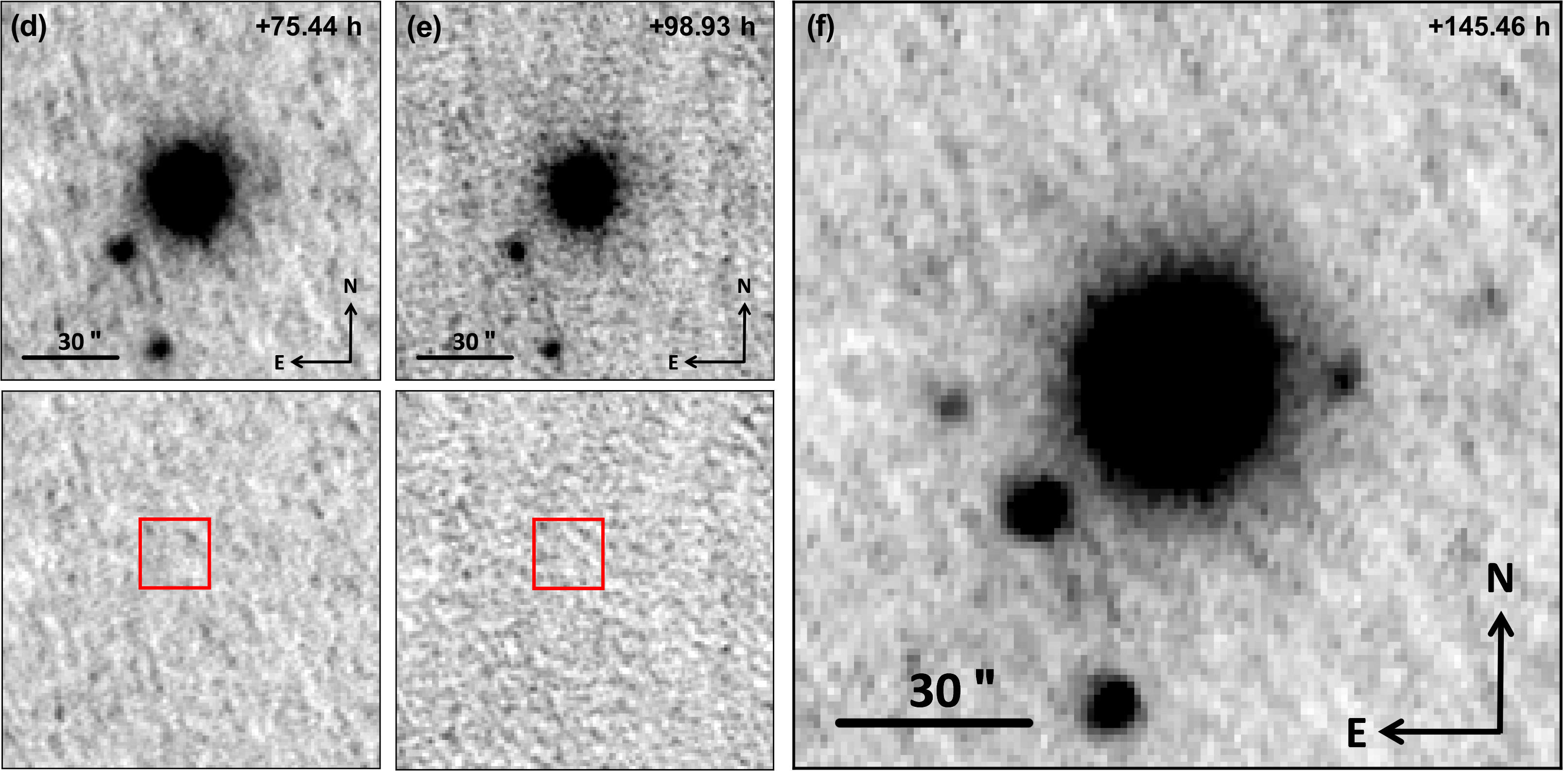}
}

\caption{$i$-band images of NGC 4993 at 5 epochs corresponding to 24.51, 25.58, 26.32, 75.44, and 98.93 hours (see Table \ref{tab1}) after LIGO trigger of GW 170817, shown in panel (a), (b), (c), (d), and (e) respectively. The upper rows of panels (a)--(e) show the co-added images of the galaxy NGC 4993, and the lower rows show the data with the template image subtracted. The red squares mark the position of the kilonova associated with GW170817. Panel (f) shows the template image used in the subtraction, constructed from observation on Aug 23, 2017 when the kilonova became too faint for detection by AST3-2.}
\label{fig2}
\end{figure*}

On Aug 23, 2017, the optical transient had faded significantly. This allowed us to use the co-added image of Aug 23, 2017 as the template for subtraction without introducing serious errors in photometric measurements of earlier data. We employed a bivariate polynomial differential background map and a spatially varying PSF for the image subtraction\citep{mill08}. Images of all other epochs were convolved/deconvolved with spatially varying PSFs to match the resolution and flux scale of the template image. The background and the shape of the PSF were solved simultaneously in Fourier space to minimize the difference between the images. After these steps, the GW optical counterpart is recovered nicely in the difference image (Figure \ref{fig2}).

The AAVSO Photometric All-Sky Survey (APASS, \cite{hen12}) was employed as the catalog for absolute flux calibration. Bright stars around NGC 4993 were selected and the optimal photometry aperture was determined using
\begin{equation}
d=2\times 0.6731\times\hbox{\rm FWHM}_{0}
\end{equation}
where $d$ is the diameter of the aperture and $\hbox{\rm FWHM}_{0}$ is the mean full-width at half-maximum of the standard stars around the target\citep{fahe14}. All photometric results are listed in Table \ref{tab1}. As shown in Figure \ref{fig2}, the optical signals associated with GW 170817 were detected at $\alpha(J2000.0)\,=\,13^{h}09^{m}48^{s}.082$, $\delta(J2000.0)\,=\,-23^{\circ}22^{'}53^{''}.60 $ on the co-added images of Aug. 18, 2017, while no detectable signal was recovered at this coordinate on Aug. 20, 2017 and Aug. 21, 2017, leading only to photometric upper limits.

\section{Theoretical Interpretation}
\begin{figure*}
\centering
\includegraphics[scale=0.54]{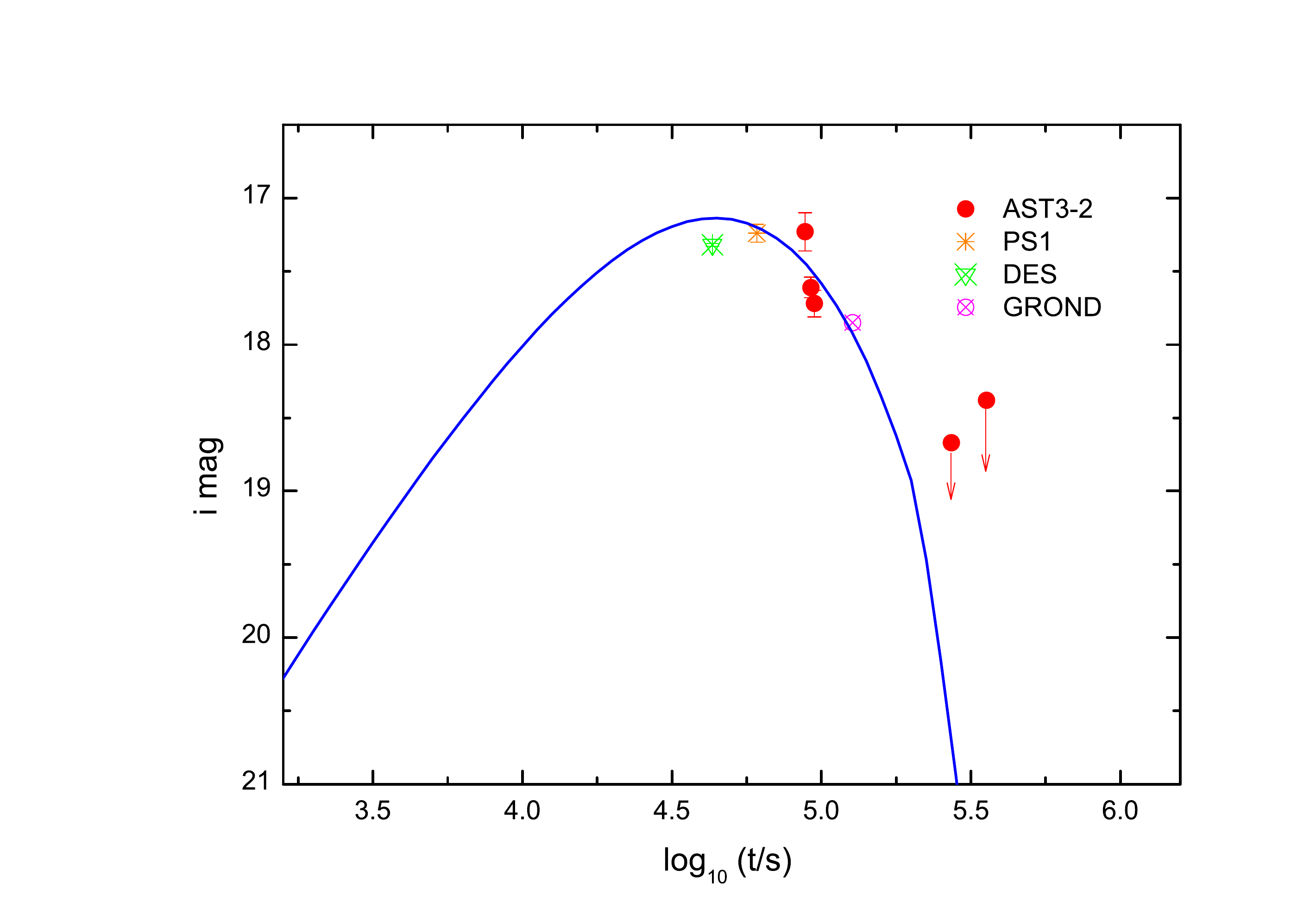}
\caption{Comparison of the analytical kilonova model with AST3-2 observational data. The X-axis is the time in seconds after the trigger of GW 170817 by LIGO and Virgo, and the Y-axis is the magnitude in $i$-band. The solid red dots are the data points from AST3-2. The data points with downward arrows are 3-$\sigma$ upper limits. Data points from DECam \citep{all17}, PS1 \citep{cham17}, and GROND \citep{smar17} are also shown. The blue solid line is the theoretical kilonova light curve, with $\kappa=10$ cm$^{2}$ g$^{-1}$, $v_{\rm{ej}}=0.29c$, and  $M_{\rm{ej}}=0.0105$ $M_{\odot}$.
\label{I-band}}
\end{figure*}

As can be seen in Table \ref{tab1} and Figure \ref{I-band}, the optical transient in $i$-band faded rapidly with a change of $\Delta m\sim 0.5$ in $\Delta t\sim 1.8$ hour at $t\sim$ 1 day after the detection of GW 170817. Such fast intra-day evolution of the light curve is consistent with the predictions of kilonova models, as also suggested by a broad range of other independent electromagnetic observations using many different telescopes by the LIGO/Virgo scientific collaboration \citep{lvc17g}.  

A general approach in determining the properties of a kilonova is to model the observed data with a detailed radiative transfer simulation \citep{met10,kas13,bar13,tan13,bar16}. However, a simple analytic model is sufficient for deriving the basic physical parameters of the kilonova. In the following, we will use a simplified kilonova model to fit the data points obtained by AST3-2.

We model the bolometric light curve of the kilonova using the formula \citep{kaw16,die17,xiao17}
\begin{equation}
L_{\text{MN}}= M_{\rm ej} \epsilon_{\rm th} \dot{\epsilon_0}  \times \left\{
\begin{array}{ll}
\frac{t}{t_c} \left(\frac{t}{1 \  \rm{day}}\right)^{-1.3}, & {\rm if}\,\,t\leq t_{c}, \\
 \left(\frac{t}{1 \  \rm{day}}\right)^{-1.3}, & {\rm if}\,\,t>t_{c},%
\end{array}%
\right.
\end{equation}
here we adopt $\dot{\epsilon}_0=1.58 \times 10^{10} {\rm erg\,g^{-1}\,s^{-1}}$ and the thermalization efficiency $\epsilon_{\rm th}=0.5$ following \citep{die17}. $t_{c}$ is the critical time, after which ($t>t_{c}$) the ejecta becomes transparent. \cite{kaw16} and \cite{die17} found that the bolometric light curve of this analytic model function can reasonably match the results of the radiation-transfer simulation performed in \cite{tan13}.

The spectrum of the kilonova emission is determined by complex frequency-dependent radiative processes within its ejecta. For simplicity, we assume that the radiation can be approximated by a blackbody spectrum. The effective temperature of the photosphere can be written as
\begin{equation}
T_{\text{eff}}=\left( \frac{L_{\text{MN}}}{\sigma
_{\text{\tiny SB}}S}\right) ^{1/4},
\end{equation}%
where $\sigma_{\rm\tiny SB}$ is the Stephan-Boltzmann constant and the emitting area  $S=4\pi R_{\rm ej}^2$
with $R_{\rm ej}\simeq v_{\text{ej}}t$ being the radius of the ejecta. 

The redshift of the host galaxy NGC 4993 is $z=0.009727$, which corresponds to a luminosity distance of $D_{L}=40$ Mpc adopting recent cosmological model parameters \citep{freedman17}. Figure~\ref{I-band} shows a preliminary fit to the AST3-2 data with our analytic kilonova model. The fitting parameters are: the opacity of the ejecta $\kappa=10$ cm$^{2}$ g$^{-1}$, the ejecta velocity $v_{\rm{ej}}=0.29c$ ($c$ is the speed of light), and the mass of the ejecta $M_{\rm{ej}}=0.0105$ $M_{\odot}$. These numbers are consistent with typical values obtained by numerical simulations of BNS mergers \citep{die17}. Note that the opacity of the r-process ejecta with the lanthanides, $\kappa=10 - 100$ cm$^{2}$ g$^{-1}$, is much higher than the opacity for iron-peak elements. The above physical parameters of the ejecta for the kilonova associated with GW 170817 can be accurately determined if other multi-wavelength observations are combined with the AST3-2 data.

\section{Summary}

GW 170817 is the first binary neutron star merger system detected by the LIGO and Virgo detector network via gravitational wave detection. Less than a few seconds after the GW detection, a short gamma ray burst GRB 170817A, was detected by the \emph{Fermi} and \emph{INTEGRAL} satellites \citep{connaughton17,savchenko17}. The optical counterpart AT2017gfo associated with GW 170817 was discovered by the 1-m Swope telescope \citep{coulter17} and independently by several other telescopes \citep{lvc17g}. The host galaxy of this counterpart is NGC 4993, with a redshift of $z=0.009727$ and corresponding distance from the Earth of $D_{L}=40$ Mpc.

The AST3-2 data revealed a fast evolving transient at $t\sim 1$ day after the GW trigger, with the $i$-band magnitude fading from $17.23\pm0.13$ magnitude at $t=24.51$ hours to $17.72\pm0.09$ magnitude at $t=26.32$ hours. We also obtained two upper limits on the brightness of the optical counterpart at later times. 

The brightness and temporal evolution of the detected optical transient associated with GW 170817 is in agreement with the predictions of simple kilonovae/macronovae models. Our observations support binary neutron star merger models for GW 170817. We also performed a model fit to the AST3-2 data, which leads to a preliminary estimate of the ejecta parameters. In this model, about $\sim 10^{-2}$ solar mass of radioactive material was ejected during the merger up to velocities as high as $30\%$ the speed of light.

\section{Acknowledgements}

The AST3 project is supported by the National Basic Research Program (973 Program) of China (Grant No.\ 2013CB834900), and the Chinese Polar Environment Comprehensive Investigation $\&$ Assessment Program (Grant No.\ CHINARE2016-02-03), the National Natural Science Foundation of China (Grant Nos.\ 11573014, 11673068, 11325313, 11633002, and 11433009), and the Key Research Program of Frontier Sciences (QYZDY-SSW-SLH010, QYZDB-SSW-SYS005), the Strategic Priority Research Program Multi-waveband gravitational wave Universe (Grant No.\ XDB23040000) and the Youth Innovation Promotion Association (Gratn No.\ 2011231) of Chinese Academy of Sciences. The construction of the AST3 telescopes was made possible by funds from Tsinghua University, Nanjing University, Beijing Normal University, University of New South Wales, Texas A$\&$M University, the Australian Antarctic Division, and the National Collaborative Research Infrastructure Strategy (NCRIS) of Australia. It has also received funding from the Chinese Academy of Sciences through the Center for Astronomical Mega-Science and National Astronomical Observatory of China (NAOC). This research was made possible through the use of the AAVSO Photometric All-Sky Survey (APASS), funded by the Robert Martin Ayers Sciences Fund. Part of this research was funded by the Australian Research Council (ARC) Centre of Excellence for Gravitational Wave Discovery (OzGrav), CE170100004, the ARC Centre of Excellence for All-sky Astrophysics (CAASTRO), CE110001020, and the Centre of Excellence for All-sky Astrophysics in 3-Dimensions (ASTRO-3D), CE170100013.  Research support to IA is provided by the Australian Astronomical Observatory (AAO). JC acknowledges the ARC Future Fellowship grant, FT130101219.

\clearpage

\clearpage

\begin{table}
\caption{AST3-2 Photometric Data of the Kilonovae Associated with GW170817}
\begin{center}
\begin{footnotesize}
\begin{tabular}{lccccccc}
\hline
IMAGE ID&UT obs start&UT obs end&UT obs average&$m_{i}$&$\Delta m_{i}$ &3 $\sigma$ $m_{i,\rm{limit}}$&Caveat\\
 &  &  &  &(mag)&(mag)&(mag)&  \\
\hline
0818a & 2017 08 18.54773909 & 2017 08 18.55200144 & 2017 08 18.54987027 & 17.23 & 0.13 & 18.25 & ...\\
0818b & 2017 08 18.58207886 & 2017 08 18.60667793 & 2017 08 18.59437840 & 17.61 & 0.07 & 19.47 & ...\\
0818c & 2017 08 18.60671538 & 2017 08 18.64366066 & 2017 08 18.62518802 & 17.72 & 0.09 & 19.53 & ...\\
0820  & 2017 08 20.59673272 & 2017 08 20.74696424 & 2017 08 20.67184848 & 18.67 & ... & 18.67 & Upper Limit\\
0821  & 2017 08 21.54887044 & 2017 08 21.75227890 & 2017 08 21.65057467 & 18.38 & ... & 18.38 & Upper Limit\\
0823  & 2017 08 23.55072740 & 2017 08 23.62820195 & 2017 08 23.58946468 & ... & ... & 19.64 & Template Image\\
\hline
\end{tabular}
\label{tab1}
\end{footnotesize}
\end{center}
\footnotesize
{Left to right columns are: 
(1) coadd image ID,
(2) start time of observation for coadded image,
(3) end time of observation for coadded image,
(4) average time of the observation,
(5) $i$-band magnitude,
(6) photometric error of $i$-band magnitude,
(7) $3\sigma$ limiting magnitude of the coadded image, and
(8) caveat on the coadded image}
\end{table}
\end{document}